\documentclass{jaa}
\usepackage{natbib}
\usepackage{multicol}
\usepackage{graphicx}
\usepackage{multirow}
\usepackage{subfigure}
\usepackage{xcolor}
\usepackage{adjustbox}
\usepackage[colorlinks=true,linkcolor=blue,citecolor=blue]{hyperref}
\begin{document}
%\sloppy

%%paper title
%%For line breaks \\ can be used within title
\title{Detection of interstellar cyanamide (NH$_{2}$CN) towards the hot molecular core G10.47+0.03}

%%author names are separated by comma (,)
%%use \and before the last author name
%%use a * along with the number separated by comma
%% for the  author for correspondence
%%\textsuperscript{number} is used for affiliation
%%\affilOne, \affilTwo etc., upto \affilTwentyfive is possible
%%Please note the first letter after \affil is capitalised in the command
%%

\author{Arijit Manna\textsuperscript{1}, Sabyasachi Pal\textsuperscript{1,*}}
\affilOne{\textsuperscript{1}Midnapore City College, Kuturia, Bhadutala, Paschim Medinipur, West Bengal, India 721129\\}

%%escape two-column mode for the title, affiliation, and abstract
%%by giving \twocolumn command as shown
\twocolumn[{
\maketitle
%%include \corres to print the corresponding author Email id
\corres{sabya.pal@gmail.com}

%%include \msinfo for
%%manuscript information such as
%received, revised, and accepted dates
%%
\msinfo{09 Decembar 2021}{24 May 2022}
%%abstract

\vspace{0.5cm}
\begin{abstract}
In the interstellar medium, the amide-type molecules play an important role in the formation of the prebiotic molecules in the hot molecular cores or high mass star formation regions. The complex amide-related molecule cyanamide (NH$_{2}$CN) is known as one of the rare interstellar molecule which has a major role in the formation of urea ({NH$_{2}$CONH$_{2}$}). In this article, we presented the detection of the emission lines of cyanamide (NH$_{2}$CN) towards the hot molecular core G10.47+0.03 between the frequency range 158.49--160.11 GHz using the Atacama Large Millimeter/submillimeter Array (ALMA) interferometric radio telescope. The estimated column density of the emission lines of NH$_{2}$CN using the rotational diagram model was $N$(NH$_{2}$CN) = (6.60$\pm$0.1)$\times$10$^{15}$ cm$^{-2}$ with rotational temperature ($T_{rot}$) = 201.2$\pm$3.3 K. The fractional abundance of NH$_{2}$CN with respect to H$_{2}$ towards the G10.47+0.03 was $f$(NH$_{2}$CN) = 5.076$\times$10$^{-8}$. Additionally, we estimated the NH$_{2}$CN/NH$_{2}$CHO abundance ratio towards the G10.47+0.03 was 0.170, which was nearly similar with NH$_{2}$CN/NH$_{2}$CHO abundance ratio towards the IRAS 16293--2422 B and Sgr B2 (M). We found that the observed abundance of NH$_{2}$CN with respect to H$_{2}$ towards the G10.47+0.03 fairly agrees with the theoretical value predicted by \citet{gar13}. We also discussed the possible formation and destruction pathways of NH$_{2}$CN.
\end{abstract}

%%insert keywords separated by 3 hyphens using \keywords{words}
%%close the twocolumn escape here
%%include \doinum{number}for the DOI number in the header
%%include \volnum{number} for the volume number in the header
%%include \year{yyyy} for  year of publication in the header
%%include \pgrange{num--num} page range of article in the header
%%include \artcitid{num} for the article citation id
%%include \lp to print last page of the article
%%include \setcounter{page}{pagenum} for the exact starting page of the article
}]
\doinum{xyz/123}
\artcitid{\#\#\#\#}
\volnum{000}
\year{2021}
\pgrange{1--11}
\setcounter{page}{1}
\lp{11}

\section{Introduction}
\label{sec:intro} 
In the interstellar medium (ISM), cyanamide (NH$_{2}$CN) is one of the rare amide-type molecule that was created with two individual nitrogen atoms. In prebiotic chemistry, the complex organic molecule NH$_{2}$CN is converted into urea (NH$_{2}$CONH$_{2}$) after the reaction with H$_{2}$O, which plays a very important role in the formation of the biochemical and biological process \citep{kil47}. The NH$_{2}$CN was the first interstellar molecule that contained an NCN frame \citep{tur75}. The NH$_{2}$CN molecule has a pyramidal equilibrium structure with 0$^{+}$ and 0$^{-}$ substates \citep{sharma21}. The electric dipole moments of NH$_{2}$CN molecule with 0$^{+}$ and 0$^{-}$ substate were $\mu_{a}=4.25\pm0.02$ D and $\mu_{a}=4.24\pm0.02$ D, respectively \citep{bro85}. In ISM, carbodiimide isomer (HNCNH) was created due to the photochemical reaction of NH$_{2}$CN and induced thermal reactions in the presence of interstellar ice analogs \citep{duv05}. The carbodiimide molecule was used to link biological processes with the assembly of peptides and amino acids \citep{wil81}.  Earlier, the emission lines of NH$_{2}$CN were detected towards the hot molecular core (HMC) Sgr B2 \citep{tur75}, Orion KL \citep{whi03}, and the high-mass protostar IRAS 20126+410 \citep{pal17}. The emission lines of NH$_{2}$CN were tentatively detected towards the low-mass star formation region Barnard 1b \citep{mar18}. Recently, an NH$_{2}$CN molecule was detected towards the low mass solar-type protostar IRAS 16293--2422 B using ALMA in the framework of the Protostellar Interferometric Line Survey (PILS) \citep{cou18}. Towards the IRAS 16293--2422 A, \citet{cou18} do not see any evidence of NH$_{2}$CN because the emission lines are broader than 16293--2422 B ($\leq$2 km s$^{-1}$). The emission lines of NH$_{2}$CN was also detected from NGC 1333 IRAS2A using the Plateau de Bure Interferometer (PdBI) \citep{cou18}. Outside of Milkyway, NH$_{2}$CN molecule was also detected towards the galaxy NGC 253 and M82 \citep{mar06,ala11}.

\begin{figure}
	\centering
	\includegraphics[width=0.43\textwidth]{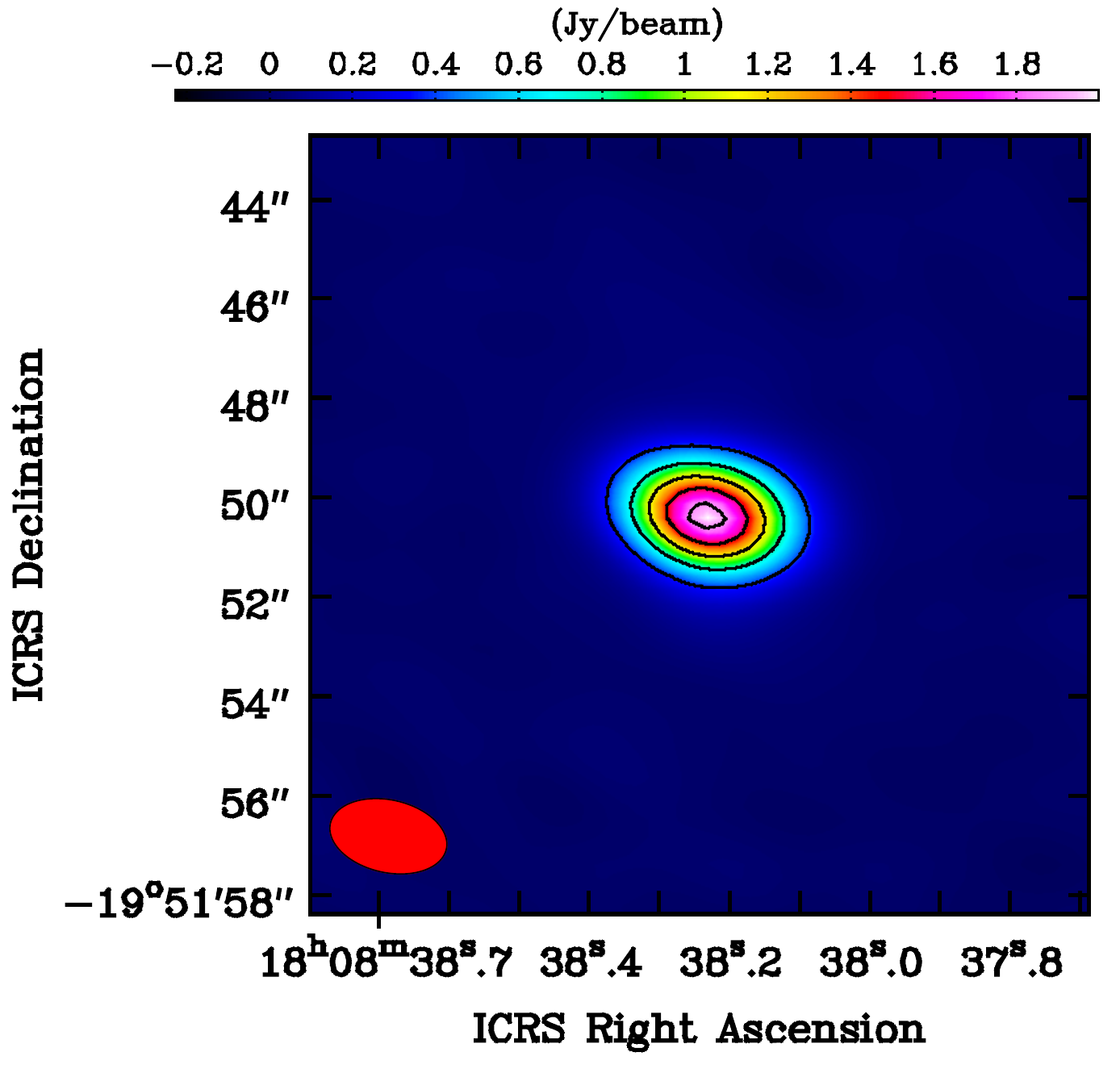}
	\caption{Millimeter continuum image of G10 obtained with ALMA at frequency 159.92 GHz. The contour level started at 3$\sigma$ where 1$\sigma$ = 9.02 mJy/beam and increasing by a factor of $\surd$2. The red color circle indicated the synthesized beam of the continuum image.}
	\label{fig:cont}
\end{figure}

\begin{table*}{}
	%	\begin{minipage}[t]{\columnwidth}
		\centering
%	\scriptsize
	\caption{Observation summary of G10}
%	\begin{adjustbox}{width=1.01\textwidth}
		\begin{tabular}{|c|c|c|c|c|c|c|c|c|c|c|c|}
		\hline
Observation date &	On-source time&Frequency range&Spectral resolution &Sensitivity (10 Km s$^{-1}$)\\	
(yyyy-dd-mm)     &	(hh:mm)       &(GHz)            & (kHz)                 &(mJy beam$^{-1}$)\\
\hline		
2017-07-03       &	00:28.72      &158.49--159.43 &1128.91               &	0.78 	\\
 --              &   --           &159.18--159.65 &~488.28                            &0.80\\
 --              &   --           &159.49--160.43 &1128.91   &0.79\\
--               &   --           &159.64--160.11 &~488.28     &0.79\\

			\hline
		\end{tabular}	
	%\end{adjustbox}

	\label{tab:data}
	%	\end{minipage}[t]{\columnwidth}
\end{table*}

In ISM, G10.47+0.03 (hereafter, G10) is a highly chemically rich HMC that was located at a distance of 8.6 kpc \citep{san14}. The luminosity of G10 was $\sim$10$^{6}$ L$_{\odot}$, which is one of the highest luminosity star-forming region in the galaxy and particular interest to the investigation of the complex molecular lines \citep{rol09}. This HMC is one of the young cluster in ISM. Earlier, \citet{gor20} detected the emission lines of formamide (NH$_{2}$CHO), methyl isocyanate (CH$_{3}$NCO), and isocyanic acid (HNCO) towards the G10 using ALMA. Recently, the emission lines of aldehyde and alcohol-related molecules, such as methanol (CH$_{3}$OH), ethylene glycol ((CH$_{2}$OH)$_{2}$), acetaldehyde (CH$_{3}$CHO), propanal (CH$_{3}$CH$_{2}$CHO), glycolaldehyde (HOCH$_{2}$CHO), acetone (CH$_{3}$COCH$_{3}$), methyl formate (CH$_{3}$OCHO), and dimethyl ether (CH$_{3}$OCH$_{3}$) were also detected towards the G10 using ALMA \citep{mondal21}. The complex prebiotic molecule methenamine (CH$_{2}$NH) and methylamine (CH$_{3}$NH$_{3}$) were also detected towards the G10, which was known as a possible precursor of the simplest amino acid glycine \citep{suz16, ohi19}.  Recently, the emission lines of another possible glycine precursor molecule, amino acetonitrile (NH$_{2}$CH$_{2}$CN) were also detected towards the G10 using ALMA \citep{man22}. After the successful detection of amino acetonitrile towards the G10, \citet{man22} also estimated the upper limit column density of glycine conformer I and II was $\leq$1.25$\times$10$^{15}$ cm$^{-2}$ and $\leq$4.86$\times$10$^{13}$ cm$^{-2}$.  %{\color{blue}After the detection of NH$_{2}$CN towards Sgr B2 and Orion KL hot molecular cores, we first time detected the spectral line of NH$_{2}$CN with emission feature towards hot molecular core G10.47+0.03.}

In this article, we presented the interferometric detection of cyanamide (NH$_{2}$CN) towards the G10 between the frequency range 158.49--160.11 GHz using ALMA. The observations and data reduction were presented in Section~\ref{obs} The result of the detection of emission lines of NH$_{2}$CN was shown in Section~\ref{res} The discussion and conclusion were shown in Section~\ref{dis} and Section~\ref{conclu}

\begin{figure*}
	\centering
	\includegraphics[width=1.01\textwidth]{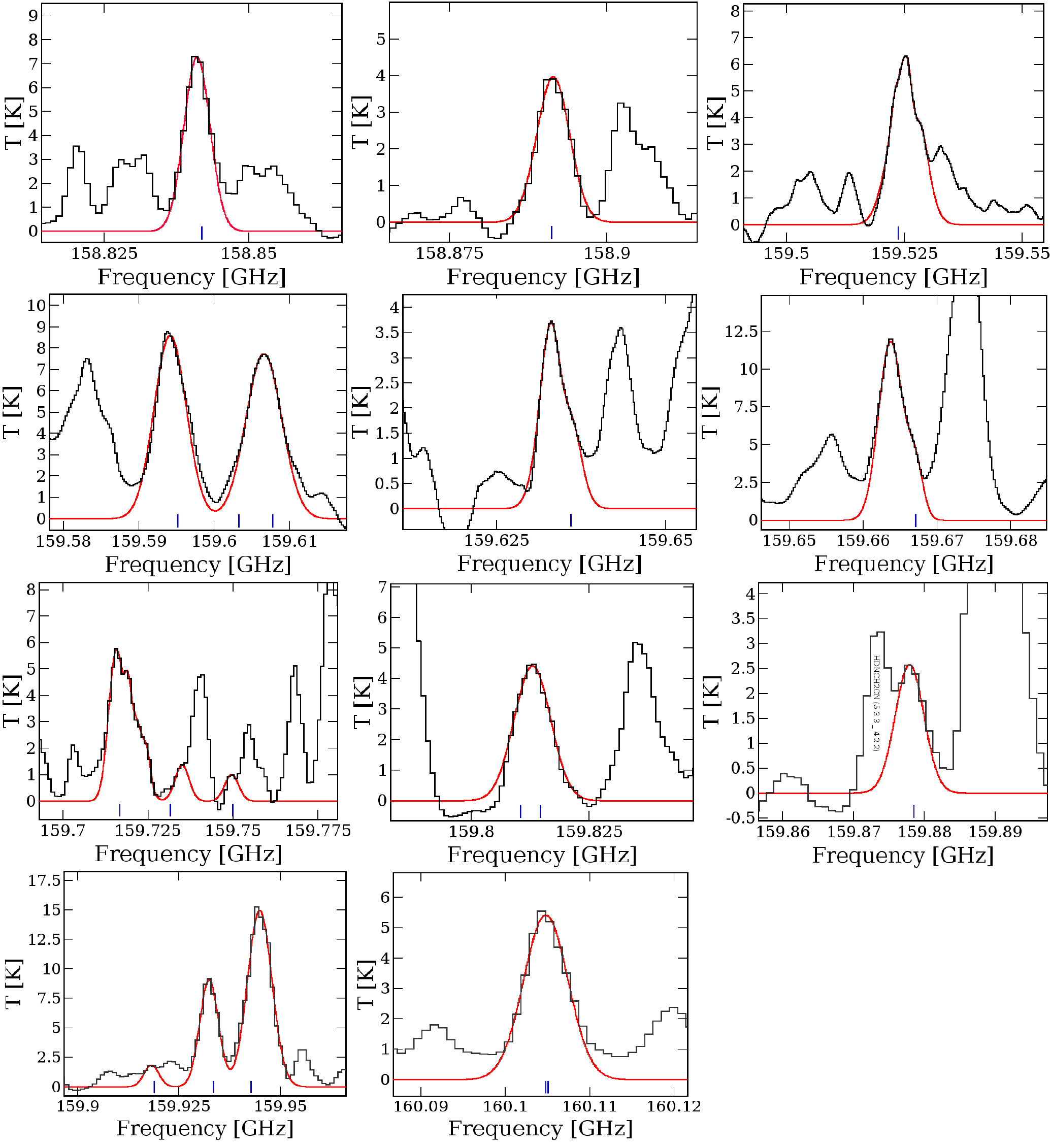}
	\caption{Rotational emission lines of NH$_{2}$CN with different transitions towards the G10. The black line showed the observed millimeter spectra, while the red synthetic line showed the best fitting Gaussian model of NH$_{2}$CN. The blue lines indicated the peak position of the detected transitions of NH$_{2}$CN.}
	\label{fig:emission} 
\end{figure*}

\section{Observations and data reductions}
\label{obs}
  To search for the emission lines of complex organic molecular gases towards the HMC candidate G10, we used the interferometric raw data of Atacama Large Millimeter/Submillimeter Array (ALMA)\footnote{\href{https://almascience.nao.ac.jp/asax/}{https://almascience.nao.ac.jp/asax/}} cycle 4 with 12 m array data (\#2016.1.00929.S., PI: Ohishi, Masatoshi). The observation of molecular lines towards the G10 was carried out with ALMA band 4 with an observed phase center ($\alpha,\delta$)$_{\rm J2000}$ = (18:08:38.232, --19:51:50.400). A total of thirty-nine antennas were used during the observation of G10. The observation aims to study the five-strong emission lines of the simplest amino acid glycine towards the G10. The observed frequency range was taken between 158.49--160.43 GHz with an angular resolution of 1.76$^{\prime\prime}$. During the observation of G10, the flux calibrator was taken as J1733--1304, the phase calibrator was taken as J1832--2039, and the bandpass calibrator was taken as J1924--2914. The systematic velocity ($V_{LSR}$) of G10 was 67--68 km s$^{-1}$ \citep{rof11}. During the observation, the minimum baseline was 15 m and the maximum baseline was 331 m. The observation summary is presented in Table~\ref{tab:data}. \cite{gor20} presented a detailed description of the observations of G10 using ALMA. 

\begin{table*}
	%	\begin{minipage}[t]{\columnwidth}
	\centering
	%	\scriptsize 
	\caption{Summary of the line properties of the NH$_{2}$CN  towards the G10.}
	\begin{adjustbox}{width=0.95\textwidth}
		\begin{tabular}{ccccccccccccccccc}
			\hline 
Observed frequency &Transition & $E_{u}$ & $A_{ij}$ &Peak intensity&S$\mu^{2}$&FWHM & V$_{LSR}$ & $\rm{\int T_{mb}dV}$ &Remark\\
(GHz) &(${\rm J^{'}_{K_a^{'}K_c^{'}}}$--${\rm J^{''}_{K_a^{''}K_c^{''}}}$) &(K)&(s$^{-1}$) &(K)&(Debye$^{2}$)& (km s$^{-1}$) &(km s$^{-1}$)&(K km s$^{-1}$) & \\
			\hline
158.8418&8(1,8)--7(1,7),V = 1&119.75&3.95$\times$10$^{-4}$ &~~7.297&143.888&10.585$\pm$0.46&68.701$\pm$0.14&~~~~68.202$\pm$5.69&Non blended\\	

158.8911&8(1,8)--7(1,7), V = 0&~~48.81&4.05$\times$10$^{-4}$&~~3.902&442.242&10.242$\pm$0.42&68.683$\pm$0.15&~~~~42.971$\pm$3.22&Non blended\\

~159.5236$^{*}$&8(7,1)--7(7,0), V = 1$^{\bullet}$&795.05&9.50$\times$10$^{-5}$&~~6.298&~~34.195&10.138$\pm$0.37&68.539$\pm$0.72&~~~~52.203$\pm$6.75&Slightly blended\\

159.5951&8(2,7)--7(2,6), V = 1 &162.29&3.73$\times$10$^{-4}$&~~8.726&401.860&10.748$\pm$0.38&68.635$\pm$0.93&~~70.277$\pm$11.26&Non blended\\

~159.6032$^{*}$&8(6,2)--7(6,1), V = 1&612.81&1.78$\times$10$^{-4}$&~~7.650&191.434&10.835$\pm$0.82&68.236$\pm$0.21&~~83.625$\pm$16.36&Blended \\

159.6077&8(2,6)--7(2,5), V = 1 &162.30&3.73$\times$10$^{-4}$&~~7.650&401.798&10.895$\pm$0.83&68.238$\pm$0.23&~~83.869$\pm$16.27&Blended\\

~159.6360$^{*}$&8(7,1)--7(7,0), V = 0 &740.02&9.78$\times$10$^{-5}$&~~3.769&105.364&10.704$\pm$0.54&68.784$\pm$0.18&~~~~31.906$\pm$5.21&Blended \\

~159.6671$^{*}$&8(5,3)--7(5,2), V = 1 &458.26&2.48$\times$10$^{-4}$&11.912&~~88.867&10.582$\pm$1.05&68.982$\pm$0.88&~~~~26.221$\pm$2.98&Blended \\

~159.7167$^{*}$&8(4,4)--7(4,3), V = 1 &331.58&3.05$\times$10$^{-4}$&~~5.380&328.002&10.956$\pm$0.98&68.597$\pm$0.69&100.453$\pm$23.29&Blended\\

~159.7316$^{*}$&8(6,2)--7(6,1), V = 0 &553.82&1.83$\times$10$^{-4}$&~~0.613&~~65.570&10.236$\pm$0.76&68.693$\pm$0.82&~~~~~~1.696$\pm$0.19&Blended\\

159.7499&8(3,6)--7(3,5), V = 1 &232.90&3.49$\times$10$^{-4}$&~~0.826&125.180&10.683$\pm$0.84&68.726$\pm$0.65&~~~~~~5.108$\pm$0.96&Blended\\

~159.8104$^{*}$&8(5,3)--7(5,2), V = 0 &395.72&2.55$\times$10$^{-4}$&~~4.469&274.038&10.490$\pm$0.64&68.597$\pm$0.76&~~63.695$\pm$12.27&Blended\\

159.8147&8(0,8)--7(0,7), V = 1 &105.83&4.08$\times$10$^{-4}$&~~4.469&438.265&10.490$\pm$0.69&68.590$\pm$0.79&~~69.428$\pm$10.38&Blended\\

~159.8785$^{*}$&8(4,4)--7(4,3), V = 0 &266.09&3.15$\times$10$^{-4}$&~~2.564&112.499&10.368$\pm$1.69&68.503$\pm$0.51&~~~~22.557$\pm$3.69&Non blended\\

159.9188&8(2,7)--7(2,6), V = 0 &~~92.51&3.93$\times$10$^{-4}$&~~1.838&140.305&10.823$\pm$0.75&68.561$\pm$0.92&~~~~18.661$\pm$1.29&Blended\\

159.9335&8(2,6)--7(2,5), V = 0 &~~92.51&3.93$\times$10$^{-4}$&~~9.869&140.313&10.826$\pm$0.79&68.596$\pm$0.73&~~85.148$\pm$10.36&Non blended\\

159.9427&8(0,8)--7(0,7), V = 0 &~~34.54&4.20$\times$10$^{-4}$&15.239&149.774&10.328$\pm$0.69&68.694$\pm$0.82&157.867$\pm$16.38&Non blended\\

160.1040&8(3,6)--7(3,5), V = 0 &164.91&3.67$\times$10$^{-4}$&~~5.569&391.627&10.068$\pm$1.92&68.793$\pm$1.29&~~57.907$\pm$10.23&Non blended\\

160.1051&8(3,5)--7(3,4), V = 0 &164.91&3.67$\times$10$^{-4}$&~~5.570&391.627&10.061$\pm$1.98&68.793$\pm$1.23&~~57.906$\pm$10.20&Non blended\\

			\hline
		\end{tabular}	
	\end{adjustbox}
	\\
	${*}$ -- Transitions contain double with frequency difference less than 100 kHz. The quantum numbers of the second are not shown.\\
	$^{\bullet}$ denoted the optical thick. 
	\label{tab:MOLECULAR DATA}
	%	\end{minipage}[t]{\columnwidth}
\end{table*}

  For the reduction of the interferometric raw data of G10, we used the Common Astronomy Software Application ({\tt CASA 5.4.1})\footnote{\href{https://casaguides.nrao.edu/}{https://casaguides.nrao.edu/}} with the standard data reduction pipeline delivered by the ALMA observatory \citep{m07}. During the analysis, the continuum flux density of the flux calibrator for each baseline was scaled and matched with the Perley-Butler 2017 flux calibrator model with 5\% accuracy using task {\tt SETJY} \citep{pa17}. The task {\tt SETJY} sets the model visibility amplitude and phase associated with a flux density scale and a specified clean component image into the model column of the data set.  After the flux, bandpass, and gain calibration using the data reduction pipeline, we separated the calibrated target data using the CASA task {\tt MSTRANSFORM} with rest frequency in each spectral window. The CASA task {\tt MSTRANSFORM} is often used after the initial calibration of the target data to make a smaller measurement set with only the data that will be used in further flagging, spectral imaging, and self-calibration. After the split of target data, We performed a first-order baseline fit to subtract the continuum emission from spectral data cubes in the CASA pipeline using the task {\tt UVCONTSUB}. For the analysis, we separated each spectral window into two data cubes, continuum, and line emission, using the task {\tt UVCONTSUB}. Now, we create the spectral data cubes of G10 using task {\tt TCLEAN} with the rest frequency of each spectral window. After the generation of the spectral data cubes, we applied the self-calibration using the tasks {\tt GAINCAL} and {\tt APPLYCAL}. After the creation of the spectral data cube, we used the task {\tt IMPBCOR} for the correction of the synthesized beam pattern. 
\begin{figure*}
	\centering
	\includegraphics[width=1.0\textwidth]{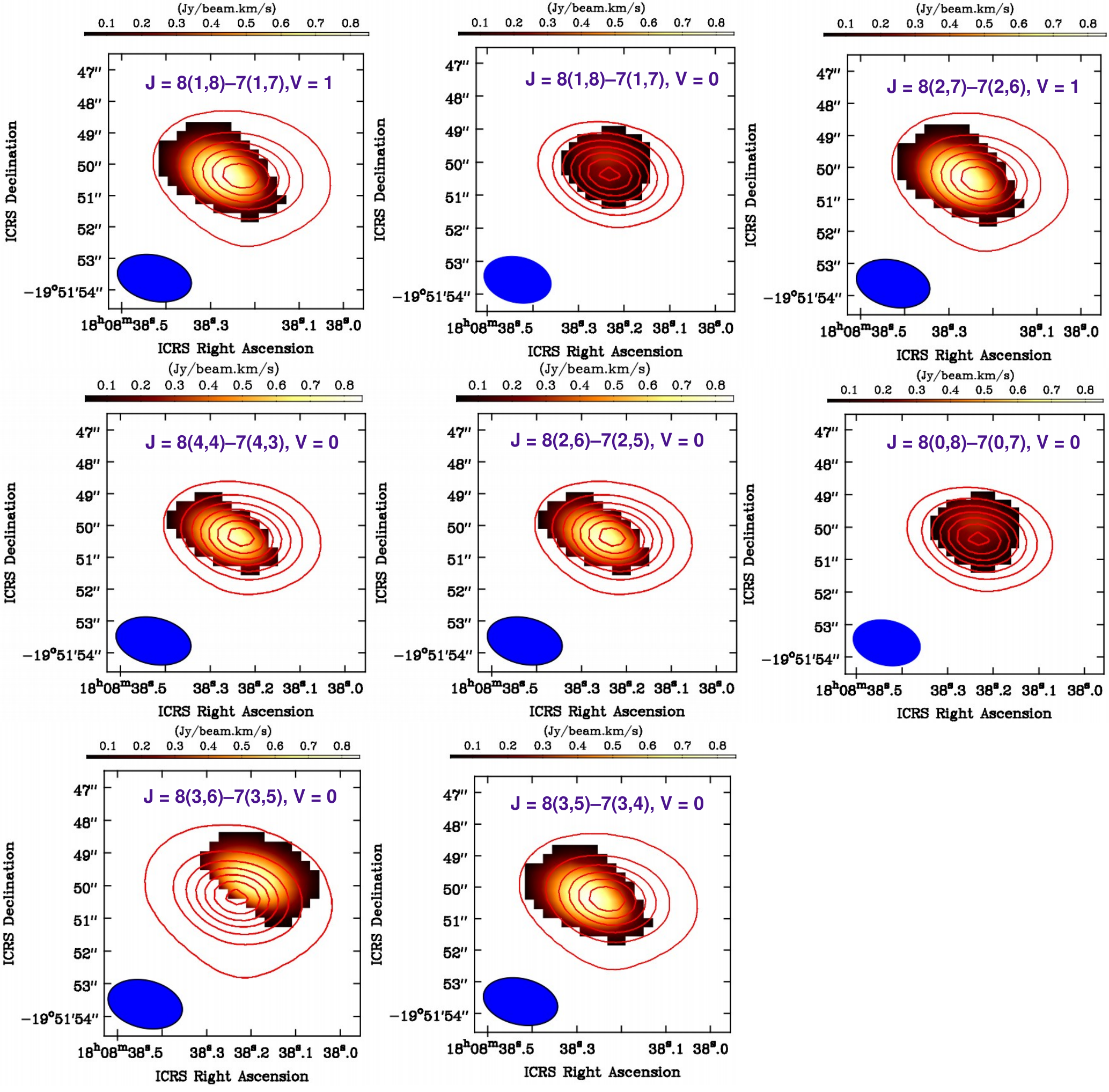}
	\caption{Integrated emission maps of unblended transitions of NH$_{2}$CN towards the G10 which was overlaid with the 1.87 mm continuum emission (red contour). The contour level started at 3$\sigma$ where 1$\sigma$ = 5 mJy/beam and increasing by a factor of $\surd$2. The blue circle indicated the synthesized beam  of integrated map.}
	\label{fig:map}
\end{figure*}

\section{Result}
\label{res}
\subsection{Continuum emission towards the G10}
We presented the millimeter continuum emission image of G10 at wavelength 1.87 mm (159.92 GHz), which is presented in Figure~\ref{fig:cont}. In a continuum image, the surface brightness colour scale has the unit of Jy beam$^{-1}$. After the generation of the continuum emission map, we fitted the 2D Gaussian over the continuum image using CASA task {\tt IMFIT} and we obtained the peak flux 1.912$\pm$0.01 Jy beam$^{-1}$ and integrated flux 2.498$\pm$0.03 Jy. The synthesized beam size of the continuum image was 2.38$^{\prime\prime}\times$1.45$^{\prime\prime}$. The position angle of the continuum image was 77.513$^{\circ}$. The observed beam size of the continuum image is not sufficient to resolve the continua. Recently, \cite{gor20} estimated the peak flux density and integrated flux density of G10 at frequency 159.45 GHz to be 1.86 Jy beam$^{-1}$ and 2.477 Jy using ALMA. Earlier, \cite{rof11} detected the millimeter continuum emission between the frequency range 201--691 GHz with the variation of flux density 6--95 Jy corresponding spectral index 2.8. 

\subsection{Line emission towards the G10}
We extracted the sub-millimeter molecular emission spectra of G10 from the continuum subtracted spectral data cubes to create a 2.5$^{\prime\prime}$ diameter circular region at the center of RA (J2000) = (18$^{h}$08$^{m}$38$^{s}$.232), Dec (J2000) =  (--19$^\circ$51$^{\prime}$50$^{\prime\prime}$.440). The systematic velocity ($V_{LSR}$) of the sub-millimeter spectra of G10 was $\sim$68.50 km s$^{-1}$. After the extraction of the chemically rich molecular sub-millimeter spectrum from G10, we used CASSIS\footnote{\href{http://cassis.irap.omp.eu/?page=cassis}{http://cassis.irap.omp.eu/?page=cassis}} \citep{vas15} for the identification of the emission lines of complex organic molecules. After deep spectral analysis, we identified the emission lines of the amide-related molecule cyanamide (NH$_{2}$CN), which was known as a precursor of urea. For identification of the emission lines of NH$_{2}$CN, we used the Cologne Database for Molecular Spectroscopy (CDMS)\footnote{\href{https://cdms.astro.uni-koeln.de/cgi-bin/cdmssearch}{https://cdms.astro.uni-koeln.de/cgi-bin/cdmssearch}} \citep{mu05} or Jet Population Laboratory (JPL) \footnote{\href{https://spec.jpl.nasa.gov/ftp/pub/catalog/catform.html}{https://spec.jpl.nasa.gov/ftp/pub/catalog/catform.html}} \citep{pic98} molecular spectroscopic databases. The hot core G10 was extremely chemically rich and the detection of the proper emission lines of NH$_{2}$CN was very difficult due to contamination of other nearby molecular transitions. We detected a total of nineteen rotational transition lines of NH$_{2}$CN between the frequency range of 158.49--160.43 GHz towards the G10. 

\begin{figure}
	\includegraphics[width=0.5\textwidth]{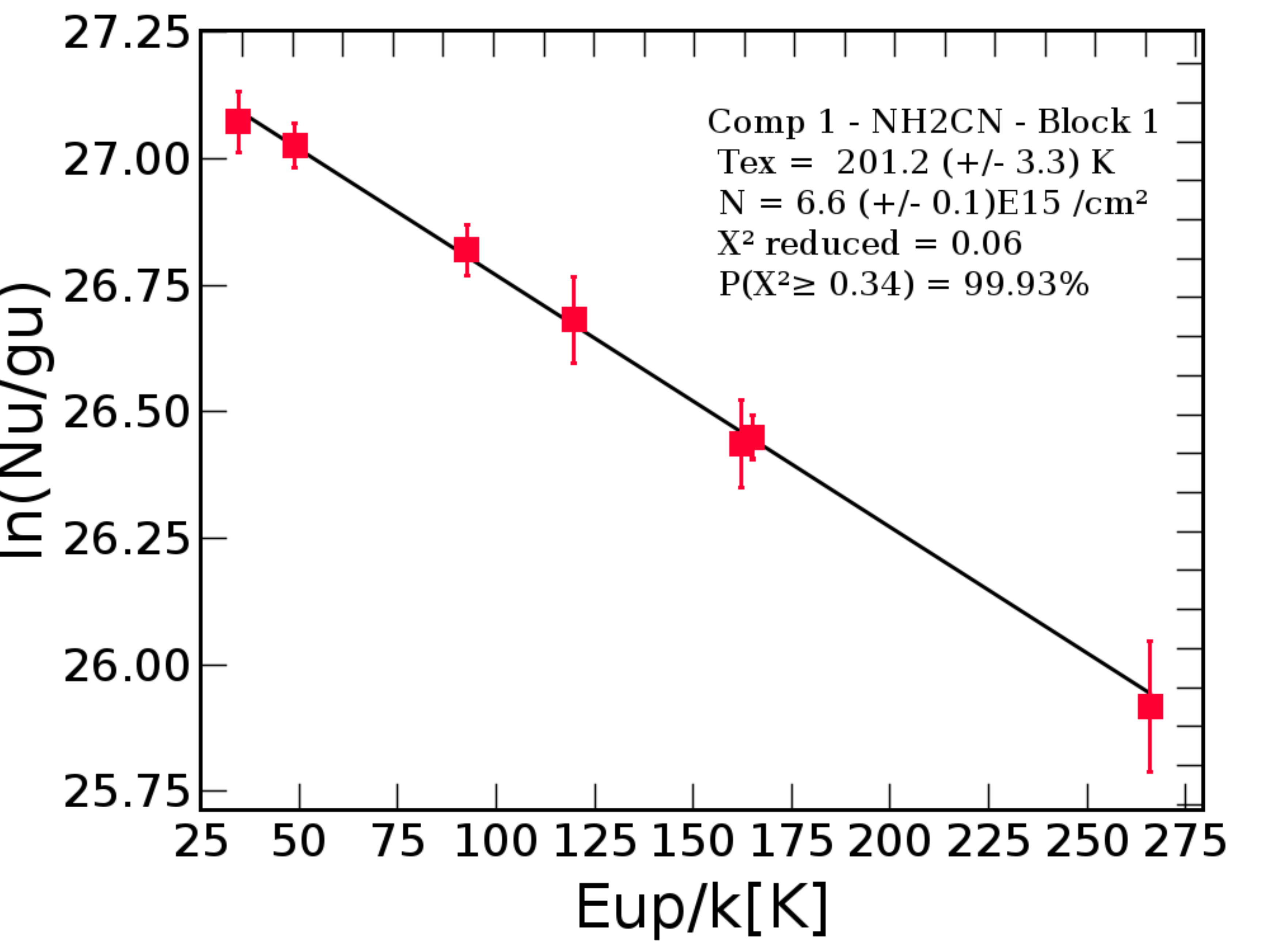}
	\caption{Rotational diagram of NH$_{2}$CN towards the G10. The red filled squares indicated the optically thin approximation data points and red  verical lines represent the error bars. The best-fit column density and rotational temperature are mentioned inside the image.}
	\label{fig:rotd} 
\end{figure}

 After the detection of the emission lines of NH$_{2}$CN from G10, we fitted the Gaussian model over the observed spectra of NH$_{2}$CN using the line analysis module in CASSIS. After the fitting of Gaussian model over the observed emission lines of NH$_{2}$CN, we calculated the Full-Width Half Maximum (FWHM), quantum numbers ({${\rm J^{'}_{K_a^{'}K_c^{'}}}$--${\rm J^{''}_{K_a^{''}K_c^{''}}}$}), upper state energy ($E_u$), Einstein coefficients ($A_{ij}$), peak intensity and integrated intensity ($\rm{\int T_{mb}dV}$). There was no missing transition of NH$_{2}$CN in this data. The summary of the detected transitions of NH$_{2}$CN was presented in Table~\ref{tab:MOLECULAR DATA} and the observed spectra of NH$_{2}$CN with Gaussian fitting were shown in Figure~\ref{fig:emission}. In Figure~\ref{fig:emission}, the black spectra indicated the observed millimeter spectra of G10, and the red synthetic spectra indicated the Gaussian model which was overlaid on the observed spectra of NH$_{2}$CN. We observed some hyperfine transitions of NH$_{2}$CN are presented corresponding to a single spectral profile because the present spectral resolution is not sufficient to resolve the emission lines of NH$_{2}$CN. The continuum emission was completely subtracted from the emission spectra. Our detected maximum transition lines of NH$_{2}$CN towards the G10 were blended with other nearby molecular lines because the observed transitions did not resolve due to the low spectral resolution. 
 	
 %	Recently, eleven unblended rotational lines of NH$_{2}$CN were detected towards the Sun-like protostar or low mass protostar IRAS 16293--2422 B, and NGC 1333 IRAS2A \citep{cou18}. Earlier, the emission lines of NH$_{2}$CN molecule were detected from the other hot molecular cores Orion KL \citep{whi03}, Sgr B2 \citep{tur75}, and IRAS 20126+4104 \citep{pal17}.

 \begin{table}{}
	%\scriptsize
	%	\begin{minipage}[t]{\columnwidth}
	\centering
	\caption{Emitting region of NH$_{2}$CN towards the G10. % which found after the best fitting of LTE model using MCMC method.
	}
	\begin{tabular}{cccccccccccc}
		\hline
		Transition&$E_{up}$&Emitting region\\
		
		(${\rm J^{'}_{K_a^{'}K_c^{'}}}$--${\rm J^{''}_{K_a^{''}K_c^{''}}}$)&(K)&[$^{\prime\prime}$]\\
		\hline
		J = 8(1,8)--7(1,7),V = 1&119.75&1.123\\
		J = 8(1,8)--7(1,7), V = 0&~~48.81 &1.139\\
		J = 8(2,7)--7(2,6), V = 1&162.29&1.196\\
		J = 8(4,4)--7(4,3), V = 0&266.09&1.263\\
		J = 8(2,6)--7(2,5), V = 0&~~92.51&1.179\\	
		J = 8(0,8)--7(0,7), V = 0&~~34.54&1.279\\
		J = 8(3,6)--7(3,5), V = 0&164.91&1.038\\
		J = 8(3,5)--7(3,4), V = 0&164.91&1.039\\

		\hline
	\end{tabular}

	\label{tab:spatial}
	%	\end{minipage}[t]{\columnwidth}
\end{table}
	
\subsection{Spatial distribution of NH$_{2}$CN}
    After the detection of the emission lines of NH$_{2}$CN towards the G10, we created the integrated emission map of NH$_{2}$CN using task {\tt IMMOMENTS} in CASA. The integrated emission map was presented in Figure~\ref{fig:map} which overlaid the 1.87 mm continuum emission. We noticed the emission map of NH$_{2}$CN has a peak at the position of the continuum. The integrated emission maps of NH$_{2}$CN were created by integrating the spectral data cubes in the velocity range where the emission lines of NH$_{2}$CN were detected. We extracted the integrated emission map only for the unblended transition of NH$_{2}$CN. The integrated emission map indicated that the different transitions of NH$_{2}$CN molecule emit from the different locations of the warm inner region of the hot core due to the different upper state energy (E$_{up}$) of all unblended transitions. We measured the emitting region of NH$_{2}$CN by fitting the 2D Gaussian of the integrated emission map of NH$_{2}$CN using CASA task {\tt IMFIT} towards the G10. The deconvolved beam size of the emitting region was calculated by the following equation\\
\begin{equation}
\theta_{S}=\sqrt{\theta^2_{50}-\theta^2_{beam}}
\end{equation}
\begin{table*}
	%	\scriptsize
	\caption{Comparison between theoritical and observed fractional abundances of NH$_{2}$CN}\label{tab:comparison} 
	%\begin{adjustbox}{width=0.7\textwidth}
	\centering      
	%\begin{center}
	\begin{tabular}{c|c|c|c|cc}
		\hline 
		& \multicolumn{3}{c}{Simulated Values$^{\rm a}$} & \multicolumn{2}{c}{Observed Values$^{\rm b}$} \\
		\hline
		Species & Fast & Medium & Slow & G10\\
		\hline
	&f(NH$_{2}$CN)~~~~~~~T (K)&f(NH$_{2}$CN)~~~~~~~T (K)&f(NH$_{2}$CN)~~~~~~T (K)& f(NH$_{2}$CN)~~~~~~~~T (K)\\
		\hline
NH$_{2}$CN		 & $2.3\times10^{-8}$~~~~~~267 & $3.6\times10^{-8}$~~~~~~~205 & $3.3\times10^{-8}$~~~~~~~195 & $5.07\times10^{-8}$~~~~~~~201.20 \\
		\hline
	\end{tabular}
	%\end{adjustbox}
	
	%\end{center}
	
	Notes: a -- Values taken from Table\,8 of \cite{gar13}; \\b --  this work.
	\label{tab:comp}
\end{table*}
where $\theta_{50} = 2\sqrt{A/\pi}$ was the diameter of the circle whose area ($A$) was enclosing $50\%$ line peak and $\theta_{beam}$ was the half-power width of the synthesized beam \citep{riv17}. The estimated emitting region of NH$_{2}$CN with unblended transitions is presented in Table~\ref{tab:spatial}. The emitting region of NH$_{2}$CN was varies between 1.038$^{\prime\prime}$--1.279$^{\prime\prime}$. We noticed that the emitting region of NH$_{2}$CN is smaller than the beam size of the integrated emission map which indicated the unblended transition lines of NH$_{2}$CN are not well spatially resolved or, at best, marginally resolved.

\subsection{Rotational diagram analysis of NH$_{2}$CN}
\label{rotd}
 We have detected the multiple hyperfine transition lines of NH$_{2}$CN towards the G10. The rotational diagram method is one of the best way to obtain the column density ($N$) in cm$^{-2}$ and rotational temperature ($T_{rot}$) in K of the detected emission lines of NH$_{2}$CN. Initially, we assumed that the detected NH$_{2}$CN spectra were optically thin and that they populated the Local Thermal Equilibrium (LTE) conditions. The assumption of LTE condition was reasonable towards the G10 because the density of inner regions of the hot core was $\sim$7$\times$10$^{7}$ cm$^{-3}$ \citep{rof11, gor20,man22}. The equation of column density for optically thin emission lines can be written \citep{gold99},

\begin{equation}
{N_u^{thin}}=\frac{3{g_u}k_B\int{T_{mb}dV}}{8\pi^{3}\nu S\mu^{2}}
\end{equation}
where, $k_B$ is the Boltzmann constant, $\rm{\int T_{mb}dV}$ is the integrated intensity, $\mu$ is the electric dipole moment, $g_u$ is the degeneracy of the upper state, $\nu$ is the rest frequency, and the strength of the transition lines were indicated by $S$. The total column density of detected species under LTE conditions can be written as,
\begin{equation}
\frac{N_u^{thin}}{g_u} = \frac{N_{total}}{Q(T_{rot})}\exp(-E_u/k_BT_{rot})
\end{equation}
 where ${Q(T_{rot})}$ is the partition function at extracted rotational temperature ($T_{rot}$) and $E_u$ is the upper state energy of the observed molecular lines. In another way, the Equation 3 can be rearranged as,
\begin{equation}
%ln\Bigg(\frac{N_u^{thin}}{g_u}\Bigg)=ln(N)-ln(Z)-(\frac{E_u}{k_B})
ln\left(\frac{N_u^{thin}}{g_u}\right) = ln(N)-ln(Q)-\left(\frac{E_u}{k_BT_{rot}}\right)
\end{equation}
Equation 4 presented a linear relationship between upper energy (E$_{u}$) and $\ln(N_{u}/g_{u}$) of the detected molecule NH$_{2}$CN. The value $\ln(N_{u}/g_{u}$) was estimated from the equation 2. From equation 4, it is indicated that spectral data points with respect to different transition lines of NH$_{2}$CN should be fitted with a straight line whose slope is inversely proportional to $T_{rot}$, with its intercept yielding $\ln(N/Q)$, which in turn will help to estimate the molecular column density. During the rotational diagram analysis, we extracted the line parameters such as FWHM, upper energy ($E_u$), Einstein coefficients (A$_{ij}$), and integrated intensity ($\int T_{mb}dV$) using a Gaussian fitting over the originally observed transition of NH$_{2}$CN using CASSIS. The fitting properties are presented in Table~\ref{tab:MOLECULAR DATA}. We noticed some hyperfine transitions of NH$_{2}$CN exists corresponding to a single spectral profile because the present spectral resolution is not sufficient to resolve the transition lines of NH$_{2}$CN. We used the high-intensity hyperfine transitions of NH$_{2}$CN in our rotational diagram analysis. We have observed the maximum transitions of NH$_{2}$CN contaminated with nearby molecular transitions. So, we used only unblended transitions of NH$_{2}$CN during the rotational diagram analysis. We noticed that the J = 8(7,1)--7(7,0), V = 1 transition line of NH$_{2}$CN was slightly blended but that line was optically thick. This transition of NH$_{2}$CN was excluded from the rotation diagram analysis because the rotational diagram method is applicable only for optically thin emission lines. The computed rotational diagram of NH$_{2}$CN was shown in Figure~\ref{fig:rotd}. In the rotational diagram, the vertical red error bars were the absolute uncertainty of $\ln(N_{u}/g_{u}$) originated from the error of the estimated $\rm{\int T_{mb}dV}$ which was calculated after fitting the Gaussian model over the detected unblended emission lines of  NH$_{2}$CN. Using the rotational diagram analysis, we estimated the column density of  NH$_{2}$CN was $N$(NH$_{2}$CN) = (6.60$\pm$0.1)$\times$10$^{15}$ cm$^{-2}$ with rotational temperature $T_{rot}$ = 201.20$\pm$3.3 K. The estimated fractional abundance of NH$_{2}$CN with respect to H$_{2}$ was $f$(NH$_{2}$CN) = 5.076$\times$10$^{-8}$ where column density of H$_{2}$ was $N$(H$_{2}$) = 1.3$\times$10$^{23}$ cm$^{-2}$ \citep{suz16}.  

We also calculated that the NH$_{2}$CN/NH$_{2}$CHO abundance ratio towards the G10 was 0.170, which agreed well with the NH$_{2}$CN/NH$_{2}$CHO ratios of the low mass star-formation region IRAS 16293 B and the hot core Sgr B2 (M). The column density of NH$_{2}$CHO towards the G10 was 3.88$\times$10$^{16}$ cm$^{-2}$ \citep{gor20}. We estimated the NH$_{2}$CN/NH$_{2}$CHO abundance ratio because NH$_{2}$CN and NH$_{2}$CHO share NH$_{2}$ as a common precursor \citep{cou18}. The NH$_{2}$CN/NH$_{2}$CHO ratio towards the IRAS 16293 B was 0.2 \citep{cou18} and Sgr B2 (M) was 0.15 \citep{bel13}. The NH$_{2}$CN/NH$_{2}$CHO ratio in another hot core Orion KL seems to be higher, i.e.$\sim$0.4--1.49 \citep{whi03}. 

\section{Discussion}
\label{dis}

\subsection{Comparison with observation and theoretical abundances of NH$_{2}$CN}
  Now we compared our calculated abundance of NH$_{2}$CN with three-phase warm-up modeling results of \citet{gar13}. \citet{gar13} assumed that there would be an isothermal collapse phase after a static warm-up phase. In the first phase, the density increases from $n_{H}$ = 3$\times$10$^{3}$ to 10$^{7}$ cm$^{-3}$ under the free-fall collapse, and the dust temperature falls to 8 K from 16 K. In the second phase, the dust temperature varies from 8 K to 400 K but the density remains constant at $\sim$10$^{7}$ cm$^{-3}$ \citep{gar13}. The number density ($n_{H}$) of G10 was $\sim$10$^{7}$ cm$^{-3}$ and temperature of the warm region was $\sim$150 K \citep{rof11, gor20, ohi19, suz16}. So, the three-phase warm-up hot core model of \citet{gar13} is suitable for understanding the chemical evolution of NH$_{2}$CN towards the G10. In the three-phase warm-up model, \citet{gar13} used the fast, medium, and slow warm-up models based on the time scale. In Table 4, we compared the estimated fractional abundance of NH$_{2}$CN with the three-phase warm-up model of \citet{gar13} and we noticed that the simulation result was nearly similar to our observational results. We found the medium warm-up model best matched the observed abundance of NH$_{2}$CN towards the G10. During the comparison of the observed abundances and simulated abundances of NH$_{2}$CN, we noticed the tolerance factor $\sim$1.5 which is a reasonably good agreement between the observational and theoretical results. Earlier, \citet{man22} and \citet{ohi19} also claimed that the medium warm-up model was the best for the environment of G10.

\subsection{Formation and destruction mechanism of NH$_{2}$CN}
 In the ISM, the low mass protostar IRAS 16293--2422 B was the only interstellar source where multiple emission lines of NH$_{2}$CN was detected \citep{cou18}. Earlier, \citet{cou18} tried to understand the formation pathways of NH$_{2}$CN towards the ISM sources without much success. According to the Kinetic Database for Astrochemistry (KIDA)\footnote{\href{http://kida.astrophy.u-bordeaux.fr/}{http://kida.astrophy.u-bordeaux.fr/}}, there are no known gas-phase reactions. In the UMIST 2012 astrochemistry chemical reaction network \citep{mce13}, there are present two known gas-phase reactions which lead to the synthesis of NH$_{2}$CN in the ISM sources. According to the UMIST 2012 astrochemistry database, the NH$_{2}$CN molecule will produce in the ISM via two pathways, (1) Neutral-neutral reaction between ammonia (NH$_{3}$) and cyanide (CN) in the grain surfaces of HMCs created NH$_{2}$CN (CN+NH$_{3}$$\longrightarrow$NH$_{2}$CN+H) \citep{smi04}, and 2) Electronic recombination of NH$_{2}$CNH$^{+}$ created NH$_{2}$CN (NH$_{2}$CNH$^{+}$+e$^{-}$$\longrightarrow$NH$_{2}$CN+H) but to create this ion is through protonation of NH$_{2}$CN itself \citep{cou18}. The reaction between NH$_{3}$ and CN is the most efficient pathway to create NH$_{2}$CN in the grain surfaces of both HMCs and solar like protostars because NH$_{3}$ and CN are the high abundant compounds in the ISM \citep{smi04, cou18}. Recently, \citet{cou18} claimed the NH$_{2}$CN molecule created towards the IRAS 16293--2422 B between the reaction of NH$_{3}$ and CN. Similarly, NH$_{2}$CN molecule will destroy via three pathways, 1) Cosmic ray-induced photo-reaction (NH$_{2}$CN+CRPHOT$\longrightarrow$NH$_{2}$+CN), 2) Ion-neutral reaction (H$_{3}$$^{+}$+NH$_{2}$CN$\longrightarrow$NH$_{2}$CNH$^{+}$+H$_{2}$), and 3) Photoprocess reaction (NH$_{2}$CN+h$\nu$$\longrightarrow$NH$_{2}$+CN). \citet{jim20} presented the possible chemical pathways to the formation of urea, 2-amino-oxazole, beta-ribocytidine-2', 3'-cyclic phosphate (pyrimidine ribonucleotide), and cytosine from NH$_{2}$CN using the biochemistry synthesis routes towards the HMC and Sun-like protostars. 

\section{Conclusion}
\label{conclu}
 In summary, we have detected evidence of the emission lines of NH$_{2}$CN which was known as a direct precursor of urea. We draw the following conclusions,\\
$\bullet$ We successfully detected a total of nineteen rotational emission lines of NH$_{2}$CN towards the hot core G10 between the frequency range of 158.49--160.11 GHz.\\
$\bullet$ We estimated the statistical column density of NH$_{2}$CN was (6.60$\pm$0.1)$\times$10$^{15}$ cm$^{-2}$ with rotational temperature 201.2$\pm$3.3 K. The fractional abundance of NH$_{2}$CN with respect to H$_{2}$ towards the G10 was 5.076$\times$10$^{-8}$. We also estimated the NH$_{2}$CN/NH$_{2}$CHO abundance ratio was 0.170.\\
$\bullet$ We encourage detailed theoretical and experimental studies using the three-phase chemical kinetics model to understand the proper formation mechanism of NH$_{2}$CN towards the HMCs, galaxies, and other chemically rich sources.

\section*{ACKNOWLEDGEMENTS}{We thank the anonymous referee for the helpful comments that improved the manuscript. The plots within this paper and other findings of this study are available from the corresponding author upon reasonable request this paper makes use of the following ALMA data: ADS /JAO.ALMA\#2016.1.00929.S. ALMA is a partnership of ESO (representing its member states), NSF (USA), and NINS (Japan), together with NRC (Canada), MOST and ASIAA (Taiwan), and KASI (Republic of Korea), in co-operation with the Republic of Chile. Joint ALMA Observatory is operated by ESO, AUI/NRAO, and NAOJ.}

\bibliographystyle{aasjournal}
%\bibliography{./literature.bib,added.bib} % if your bibtex file is called example.bib

\begin{thebibliography}{}
	\expandafter\ifx\csname natexlab\endcsname\relax\def\natexlab#1{#1}\fi
	\providecommand{\url}[1]{\href{#1}{#1}}
	\providecommand{\dodoi}[1]{doi:~\href{http://doi.org/#1}{\nolinkurl{#1}}}
	\providecommand{\doeprint}[1]{\href{http://ascl.net/#1}{\nolinkurl{http://ascl.net/#1}}}
	\providecommand{\doarXiv}[1]{\href{https://arxiv.org/abs/#1}{\nolinkurl{https://arxiv.org/abs/#1}}}
	

\bibitem[\protect\citeauthoryear{Aladro et al.}{2011}]{ala11}Aladro, R., Mart{\'\i}n, S., Mart{\'\i}n-Pintado, J. et al. 2011, A\&A, 535, A84


\bibitem[\protect\citeauthoryear{Brown et al.}{1985}]{bro85}Brown, R. D., Godfrey, P. D., Kleib{\"o}mer, B., 1985, Journal of Molecular Spectroscopy, 2, 257-273

\bibitem[\protect\citeauthoryear{Belloche et al.}{2013}]{bel13}Belloche, A., Müller, H. S., Menten, K. M., Schilke, P., \& Comito, C. 2013, A\&A, 559, A47


\bibitem[\protect\citeauthoryear{Coutens et al.}{2018}]{cou18}Coutens, A., Willis, E. R., Garrod, R. T., et al. 2018, A\&A, 612, A107

\bibitem[\protect\citeauthoryear{Duvernay et al.}{2005}]{duv05}Duvernay, F., Chiavassa, T., Borget, F., \& Aycard, J.-P. 2005, J. Phys. Chem. A, 109, 603

\bibitem[\protect\citeauthoryear{Garrod}{2013}]{gar13} Garrod, R. T. 2013, ApJ, 765, 60	

\bibitem[\protect\citeauthoryear{Gorai et al.}{2020}]{gor20}Gorai, P., Bhat, B., et al. 2020, ApJ 895 86

\bibitem[\protect\citeauthoryear{Goldsmith \& Langer}{1999}]{gold99}{Goldsmith}, P.~F., \& {Langer}, W.~D. 1999, apj, 517, 209

\bibitem[\protect\citeauthoryear{Jim{\'e}nez et al.}{2020}]{jim20}Jim{\'e}nez-Serra, I., Mart{\'\i}n-Pintado, J., Rivilla, V. M., Rodr{\'\i}guez-Almeida, L., Alonso, A. E. R., Zeng, S., Cocinero, E. J., Mart{\'\i}n, S., Requena-Torres, M., Mart{\'\i}n-Domenech, R., Testi, L.,2020, Astrobiology, 20, 9, 1048-1066	

\bibitem[\protect\citeauthoryear{Kilpatrick}{1947}]{kil47}Kilpatrick, M. L. 1947, J. Am. Chem. Soc., 69, 40

\bibitem[\protect\citeauthoryear{Mondal et al.}{2021}]{mondal21}Mondal, S. K., Gorai, P., Sil, M., Ghosh, R., Etim, E. E., Chakrabarti, S. K., Shimonishi, T., Nakatani, N., Furuya, K., Tan, J. C., Das, A., 2021, arXiv:2108.06240 


\bibitem[\protect\citeauthoryear{Mart{\'\i}n et al.}{2006}]{mar06}{Mart{\'\i}n}, S., {Mauersberger}, R., {Mart{\'\i}n-Pintado}, J., {Henkel}, C., {Garc{\'\i}a-Burillo}, S., 2006, ApJ, 164, 2,450-476


\bibitem[\protect\citeauthoryear{Marcelino et al.}{2018}]{mar18}Marcelino, N., Gerin, M., Cernicharo, J., et al. 2018, A\&A, 620, A80

\bibitem[\protect\citeauthoryear{Manna \& Pal}{2022}]{man22}Manna, A., \& Pal, S., 2022, Life Sciences in Space Research, 34, 9--15


\bibitem[\protect\citeauthoryear{M\"uller et al.}{2005}]{mu05} M\"uller, H. S. P., SchlM$\ddot{o}$der, F., Stutzki, J. \& Winnewisser, G. 2005, Journal of Molecular Structure, 742, 215--227

\bibitem[\protect\citeauthoryear{McElroy et al.}{2013}]{mce13}{McElroy}, D., {Walsh}, C., {Markwick}, A.~J., {Cordiner}, M.~A., {Smith}, K., {Millar}, T.~J., 2013, A\&A, 550, A36

\bibitem[\protect\citeauthoryear{McMullin et al.}{2007}]{m07}McMullin, J. P., Waters, B., Schiebel, D., Young, W., \& Golap, K. 2007, in Astronomical Society of the Pacific Conference Series, Vol. 376, Astronomical Data Analysis Software and Systems XVI, ed. R. A. Shaw, F. Hill, \& D. J. Bell, 127

\bibitem[\protect\citeauthoryear{Ohishi et al.}{2019}]{ohi19} Ohishi, M., Suzuki, T., Hirota, T., Saito, M. \& Kaifu, N. 2019, PASJ, 71

\bibitem[\protect\citeauthoryear{Palau et al.}{2017}]{pal17} Palau, A., Walsh, C., S{\'a}nchez-Monge, {\'A}., Girart, J. M., Cesaroni, R., Jim{\'e}nez-Serra, I., Fuente, A., Zapata, L. A., Neri, R., 2017, MNRAS, 467, 2723

\bibitem[\protect\citeauthoryear{Perley \& Butler}{2017}]{pa17}Perley, R. A., Butler, B. J. 2017, ApJ, 230, 1538

\bibitem[\protect\citeauthoryear{Pickett et al.}{1998}]{pic98}Pickett, H. M., Poynter, R. L., Cohen, E. A., et al. 1998, Journal of Quantitative Spectroscopy and Radiative Transfer, 60, 883

\bibitem[\protect\citeauthoryear{Rolffs et al.}{2011}]{rof11}Rolffs, R., Schilke, P., Zhang, Q., \& Zapata, L. 2011, A\&A, 536, A33

\bibitem[\protect\citeauthoryear{Rolffs et al.}{2009}]{rol09}Rolffs, R., Schilke, P., Zhang, Q., Wyrowski, F., Menten, K., Zapata, L., 2009, Astronomical Society of the Pacific Conference Series, 417, 215

\bibitem[\protect\citeauthoryear{Rivilla et al.}{2017}]{riv17}Rivilla, V. M., Beltran, M. T., Cesaroni, R., et al. 2017, A\&A, 598, A59

\bibitem[\protect\citeauthoryear{Smith et al.}{2004}]{smi04}Smith, I. W. M., Herbst, E., \& Chang, Q. 2004, MNRAS, 350, 323

\bibitem[\protect\citeauthoryear{Sharma}{2021}]{sharma21}Sharma, M. K., Astrophysics, 2021, 64,71-80

\bibitem[\protect\citeauthoryear{Suzuki et al.}{2016}]{suz16}Suzuki, T., Ohishi, M., Hirota, T., Saito, M., Majumdar, L., Wakelam, V., 2016, ApJ, 825, 1, 79

\bibitem[\protect\citeauthoryear{Sanna et al.}{2014}]{san14}Sanna, A., Reid, M. J., Menten, K. M., et al. 2014, ApJ, 781, 108

\bibitem[\protect\citeauthoryear{Turner et al.}{1975}]{tur75}Turner, B. E., Liszt, H. S., Kaifu, N., \& Kisliakov, A. G. 1975, ApJ, 201, L149

\bibitem[\protect\citeauthoryear{Vastel et al.}{2015}]{vas15}Vastel, C., Bottinelli, S., Caux, E., Glorian, J. M., \& Boiziot, M. 2015, in SF2A-2015: Proceedings of the Annual Meeting of the French Society of Astronomy and Astrophysics, 313--316

\bibitem[\protect\citeauthoryear{Williams \& Ibrahim}{1981}]{wil81}Williams, A., \& Ibrahim, I. T. 1981, Chem. Rev., 81, 589

\bibitem[\protect\citeauthoryear{White et al.}{2003}]{whi03}White, G. J., Araki, M., Greaves, J. S., Ohishi, M., \& Higginbottom, N. S. 2003, A\&A, 407, 589


\end{thebibliography}

%\appendix

\end{document}